\documentclass[
preprint, % preprint for 1 column, reprint for 2
notitlepage,
tightenlines, % optional: makes preprint less vertically loose
aps,
prfluids,
superscriptaddress,
longbibliography,
floatfix,
a4paper]{revtex4-2}

\usepackage{import}
\usepackage{lipsum}

\usepackage{natbib}
\usepackage{times}
\usepackage{graphicx}
\graphicspath{{fig/}}
\usepackage{amsfonts,amsmath,amsthm,amssymb,mathrsfs}
\usepackage{dsfont}
\usepackage{xcolor}
\usepackage{microtype}
\usepackage{bm}
\usepackage{siunitx}
\usepackage[hidelinks]{hyperref}

\allowdisplaybreaks

\usepackage[abs]{overpic}
\usepackage[prologue,dvipsnames]{xcolor}
\usepackage{bm}
    
\newcommand{\bk}{\bm{k}}
\newcommand{\bx}{\bm{x}}
\newcommand{\bu}{\bm{u}}
\newcommand{\bU}{\bm{U}}

\newcommand{\tih}{\tilde{h}}

\newcommand{\kU}{\bk\cdot\bU}
\newcommand{\kx}{\bk\cdot\bx}
\newcommand{\rmd}{\mathrm{d}}
\newcommand{\rme}{\mathrm{e}}
\newcommand{\rmi}{\mathrm{i}}
\newcommand{\hw}{\hat{w}}
\newcommand{\hG}{\hat{G}}
\newcommand{\hGm}{\hG_-}
\newcommand{\hGp}{\hG_+}
\newcommand{\cP}{\mathcal{P}}

\newcommand{\bcX}{\bm{\mathcal{X}}}

\newcommand{\bcF}{\boldsymbol{\mathcal{F}}}

\newcommand{\cY}{\mathcal{Y}}
\newcommand{\bcY}{\bm{\cY}}
\newcommand{\bde}{\mathbf{e}}
\newcommand{\tbcY}{\tilde{\bcY}}
\newcommand{\wint}{\sigma}%{\Omega_0}
\newcommand{\cU}{\mathcal{U}}
\newcommand{\cW}{\mathcal{W}}
\newcommand{\sT}{\mathsf{T}}
\newcommand{\Omz}{\Omega(0)}%{\Omega_0}
\newcommand{\hGw}{\hG}
\newcommand{\mA}{\bm{A}}

\newcommand{\mC}{\bm{C}}
\newcommand{\mH}{\bm{H}}

\newcommand{\mV}{\bm{V}}
\newcommand{\mR}{\bm{R}}
\newcommand{\mT}{\bm{T}}
\newcommand{\mTinf}{\mT_\infty}
\newcommand{\mCinf}{\mC_\infty}
\renewcommand{\hm}{\hat{m}}

\newcommand{\Ug}{U_\gamma}
\newcommand{\rk}{k}
\newcommand{\bsig}{\bar{\sigma}}
\newcommand{\vk}{\varkappa}
\newcommand{\vrho}{\varrho}
\newcommand{\order}[1]{\mathcal{O}(#1)}

\usepackage[capitalize]{cleveref}

\begin{document}

\title{Exact dispersion relation for linear surface waves on arbitrary vertical shear}

\author{Kjell S. Heinrich}
\author{Simen \AA. Ellingsen}
\thanks{Corresponding author: S. \AA. Ellingsen, simen.a.ellingsen@ntnu.no}
\affiliation{Department of Energy and Process Engineering, Norwegian University of Science and Technology, N-7491 Trondheim, Norway}

\begin{abstract}
      We derive the formal solution to the dispersion relation for linear surface waves on a horizontal mean current with arbitrary vertical dependence. The problem is cast in a Green’s function framework for the Rayleigh equation, neglecting viscosity but making no further approximations about the mean velocity profile. The solution is the dispersion relation in the form of a single, implicit equation relating -- and containing only -- the velocity profile, wave frequency, and wavenumber. By isolating curvature effects in a path-ordered exponential, we obtain a solution that serves as a natural starting point for systematic approximations. We demonstrate that our solution reduces to the expression found by Shrira (1993, \emph{J.\ Fluid Mech.} {\bf 252}, 565--584) in the deep-water limit, yields known asymptotic approximations, and recovers known analytical solutions in special cases.
\end{abstract}

\maketitle
%%%%%%%%%%%%%%%%%%%%%%%%%%%%%%%%%%%%%%%%%%%%%%%%%
%%%%%%%%%%%%%%%%% S E C T I O N %%%%%%%%%%%%%%%%%
%%%%%%%%%%%%%%%%%%%%%%%%%%%%%%%%%%%%%%%%%%%%%%%%%
\section{Introduction} \label{sec:Intro}

Surface gravity waves propagating on vertically sheared currents are central to a wide range of oceanographic and engineering problems, from wave forecasting and remote sensing to air–sea interaction and coastal transport. In a range of important questions involving waves and currents together, the vertical structure of a current plays an important role, not just its depth-average or surface value. Examples where vertical shear often needs to be taken into account are ocean transport of plastics and pollutants \citep{laxague2018, Lodise2019}, wind-stress modelling \citep{Ortiz2025}, wave-breaking predictions \citep{Zippel2017}, wave statistics \citep{zheng2023}, rogue-wave formation and kinematics \citep{Li2024,Ellingsen2024} and wave forces on vessels \citep{Li2019} and structures \citep{Xin2023}. 

The problem we consider is a classical one and has been treated lucidly by \citet{peregrine1976interaction}, \citet{shrira1993surface}, and many others since. We assume waves are propagating on the surface, taken to be at $z=0$ on average. Here, $z$ is the vertical axis pointing up. The waves travel on a horizontal mean current with arbitrary vertical shear, $\bm{U}(z) = [U(z), V(z),0]$. The bottom is at a constant depth $ z =- h$. We linearise the incompressible Euler and continuity equations, along with the boundary conditions at the free surface and bottom. We may then assume solutions of Fourier form $\varphi(\bx,z,t)\sim\hat\varphi(z)\exp[\rmi\psi(\bx,t)]$, with the phase $\psi(x,t)=\kx_h-\omega\, t$. The hat represents a Fourier transformed quantity, and $\varphi$ represents a wave-induced velocity $\bu= [u,v,w]$ (the $x,y$ and $z$ components), pressure perturbation, or surface elevation $\zeta$ (where $\hat{\zeta}$ is independent of $z$). The horizontal position is $\bx_h=[x,y,0]$; $\bk=[k_x,k_y,0]$ is the wave vector, and $\omega$ is the wave frequency. Only the component of the current parallel to $\bk$ enters the problem. To simplify, we define the shorthand $\Ug\equiv\kU/k=U\cos\gamma$, so that $\gamma$ is the angle between $\bk$ and $\bU$, and $k=|\bk|$. Eliminating pressure and horizontal wave velocities, one is left with a second-order differential equation for the vertical wave velocity $\hw(z)$. This is the Rayleigh, or inviscid Orr-Sommerfeld equation, in the form 
\begin{align}\label{eq:Rayleigh_w} 
    \hat{w}'' - k^2[1+q(z)]\, \hat{w}&=0,
\end{align} \newpage
\noindent with shorthands 
\begin{equation}
    q(z)\equiv \frac{-\Ug''(z)}{k\Omega(z)}; \qquad \Omega(z)\equiv\omega-k\Ug(z),
\end{equation}
where a prime denotes a derivative with respect to $z$. Note that $\Omega(z)$ is the `locally Doppler-shifted' frequency and that the intrinsic wave frequency, i.e.\ measured in the surface-following reference system, is $\sigma(\bk)=\Omega(0)$.

We now turn to the linearised boundary conditions at the bottom and surface, which can be written \citep{shrira1993surface}
\begin{equation}\label{eq:RayleighBC}
    \hw(-h)=0; \qquad \wint^2\,\hw'(0) = [(g+\Upsilon k^2)k^2-\wint\, k\, \Ug'(0)]\,\hw(0).
\end{equation}
The gravitational acceleration is $g$, and $\Upsilon$ is the kinematic surface tension coefficient. In the absence of current, the solution is $\wint=\omega=\omega_0$ with
\begin{equation}
    \omega_0\equiv\sqrt{(gk+\Upsilon k^3)\tanh kh} .
\end{equation}
Assuming $\bk$ is given, the eigenvalue problem defined by \eqref{eq:Rayleigh_w} and \eqref{eq:RayleighBC} has $\hw(z)$ as an unknown function. There are two permitted eigenvalues for frequency, $\omega=\omega_\pm(\bk)$, corresponding to waves propagating parallel and anti-parallel to $\bk$. Thus, $\omega_\pm(-\bk)=-\omega_\mp(\bk)$, so there is only one value of $\omega$ for each wave $\bk$, with $\wint(\bk)>0$. All approaches to the linear dispersion relation -- exact, approximate, or numerical -- take this classical eigenvalue problem as their starting point.

For practical purposes, it is common to derive explicit approximate expressions, such as $\omega \approx \omega(\bk;\bU)$. This method assumes that either a suitably depth-averaged mean-current shear or curvature is small. The classical and most common asymptotic form was derived for $h=\infty$ by \citet{Stewart1974} and generalised to finite depth by \citet{skop1987approximate}. It reads ${\omega(\bk;\Ug)\approx \omega_0(k)+k\langle\Ug\rangle}$ with the second term acting as an “effective Doppler shift”: 
\begin{align} 
    \langle \Ug \rangle &= 2k\int_{-h}^0\rmd z \, \Ug(z) \frac{\cosh 2k(z+h)}{\sinh 2kh} = \Ug(0)-s(\bk)\frac{\omega_0(k)}{k};\label{eq:effDopp} \\
    s(\bk)&\equiv \frac k\omega_0\int_{-h}^0\rmd z\,\Ug'(z)\frac{\sinh2 k(z+h)}{\sinh 2kh} \overset{kh\to\infty}{\longrightarrow} \frac k\omega_0\int_{-h}^0\rmd z\,\Ug'(z)\rme^{2kz}.\label{eq:s} 
\end{align} 
The final form of \eqref{eq:effDopp} is obtained by partial integration. The quantity $s(\bk)$ is a non-dimensional number (for gravity driven waves it might be called a shear--Froude number) comparing the depth-averaged shear to the shear-free wave frequency as a measure of the effect of the shear on wave dispersion; we refer to it as the wave--shear-strength number. In the last form of \eqref{eq:s}, $\tanh kh\to 1$ is understood. \Cref{eq:effDopp} was derived under the assumption of a weak current, but \citet{ellingsen2017approximate} showed the actual sufficient criterion is $s\ll 1$. Approximations are discussed further in Section \ref{sec:weakApprox}.

Alternatively, various iterative numerical schemes have been used to solve the eigenvalue problem. The most classical approach is the piecewise-linear approximation, which approximates $\Ug(z)$ by partitioning it into a finite number of linear segments \citep{Thompson1949,zhang2005short}. The dispersion relation becomes a solution of a linear system requiring removal of a large number of spurious solutions \citep{smeltzer2017surface}. \citet{Dong2012} developed a shooting algorithm for solving the Rayleigh equation, while \citet{maxwell2020} developed a powerful path-following approach suitable when thousands of calculations with very high precision are needed. The `Direct integration method' (DIM) of \citet{li2019framework} has gained some popularity due to its rapid convergence and arbitrary accuracy.

Explicit, analytical dispersion relations in the form $\omega=\omega(\bk;\Ug)$ have been found only for very special forms of $\Ug(z)$. Beyond this, the existing solutions in practical use are either approximations valid when $\Ug$ satisfies certain $k$-dependent conditions (these are common), or require solving the eigenvalue problem numerically. Here, we instead derive a dispersion relation equation of implicit form 
\begin{equation}\label{eq:closedDisprel} 
    D(\bk,\wint;\, \bU) = 0 
\end{equation} 
which is fully self-contained and does not involve any unknown eigenfunction of $z$ and allows (numerical or in special cases, analytical) evaluation of $\wint(\bk)$. The present work obtains an exact analytical dispersion relation of this form by casting the Rayleigh problem in a Green's function framework, writing the vertical-velocity Green's function as a path-ordered exponential that isolates the curvature effect. By constructing two fundamental solutions that separately satisfy the bottom and free-surface boundary conditions and then matching them, we obtain the formal solution; an implicit one-equation dispersion relation that depends only on the prescribed shear profile. This form recovers the classical no-shear and constant-vorticity limits, connects directly to the near-potentiality approximation of \citet{ellingsen2017approximate}, and provides a natural starting point for systematic approximations and numerical evaluation.

%%%%%%%%%%%%%%%%%%%%%%%%%%%%%%%%%%%%%%%%%%%%%%%%%
%%%%%%%%%%%%%%%%% S E C T I O N %%%%%%%%%%%%%%%%%
%%%%%%%%%%%%%%%%%%%%%%%%%%%%%%%%%%%%%%%%%%%%%%%%%
\section{Approach and analytical solution}
Our starting point is the eigenvalue problem in equations \eqref{eq:Rayleigh_w}--\eqref{eq:RayleighBC}. In our derivation, we assume that the surface disturbance comes from a submerged point source. In a linearised theory, a solution for a delta-function disturbance can be integrated to obtain solutions for any distributed source, which is the idea behind Green's functions. 

\begin{figure}
    \centering
    \begin{overpic}[width=0.6\textwidth]{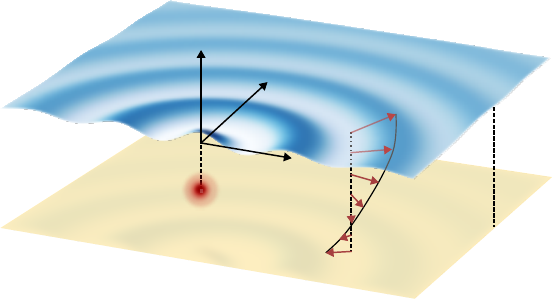}%
    \put(140,60){$x$}%
    \put(140,110){$y$}%
    \put(100,135){$z$}%
    \put(190,30){$\bm{U}(z)$}%
    \put(255,60){$h$}%
    \put(90,35){$m(t)$}%
    \put(80,60){$z_0$}%
    \end{overpic}
    \caption{Geometry of the problem set-up used for derivation.}
    \label{fig:Setup}
\end{figure}
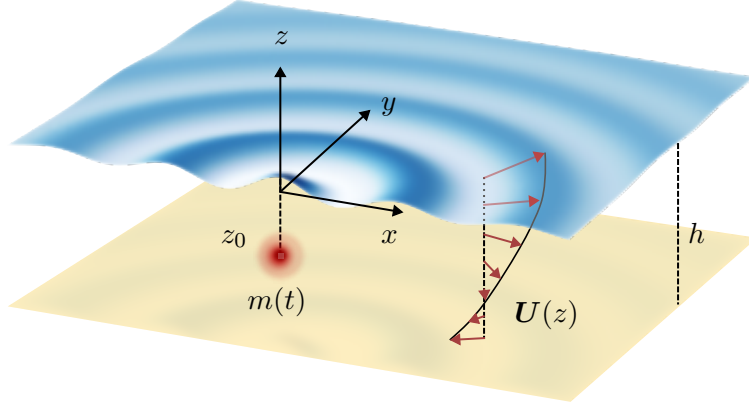

We consider the set-up shown in \cref{fig:Setup}: a point source of time-dependent strength $m(t)$ located at $\bm{x}_0=[0,0,z_0]$. The continuity equation reads
\begin{align} \label{eq:continuity}
    \nabla \cdot \bm{u}(\bm{x}) = m(t)\,\delta(\bx_h)\delta(z-z_0).
\end{align}
We assume $\bu$ and surface elevation $\zeta(\bx_h,t)$ to be small, and linearise equations and boundary conditions with respect to these. Taking the Fourier transform and combining the continuity and Euler equations in the bulk, we eliminate the horizontal velocity components and obtain 
\begin{align} \label{eq:Rayleigh_velocity_w}
    \Omega(z)\, \left[\hat{w}'' - k^2\, \hat{w} \right]\, + k\Ug''\,   
    \hat{w} = \hat{m}\left[ \Omega(z)\,\delta'(z-z_0) - k\Ug' \, \delta(z-z_0) \right].
\end{align}
We require the Fourier transformed vertical velocity $\hw$ to satisfy the boundary conditions in \eqref{eq:RayleighBC}. To eliminate $\hm$ we define a Green's function ${\hGw = \hGw(z)}$ 
(understood to depend on $\bk$ and $\omega$). We choose the Green's function to be dimensionless and required by the Rayleigh equation with a unitary source
\begin{align} \label{eq:Greens_function_definition}
     \hGw'' - k^2[1+q(z)] \hGw = \delta'(z-z_0) - (k U_\gamma'/\Omega) \, \delta (z-z_0) , 
\end{align}
with the shorthand 
\begin{equation}\label{eq:q}
    q(z) \equiv \frac{-\Ug''(z) }{k\Omega(z)}.
\end{equation}
At the source, the Green's function must obey the jump and continuity conditions. These are found by integrating the continuity equation and \eqref{eq:Greens_function_definition} just across $z_0$, and give
\begin{align} \label{eq:Bs_at_source}
    [\hGw]_{z_0} = 1,\qquad [\Omega \hGw' - \Omega'\hGw]_{z_0} = 0,
\end{align}
where $[f]_{z_0} = \lim_{\epsilon\to0^+} f(z_0 + \epsilon) - f(z_0 - \epsilon)$.  

It is not possible to obtain a closed-form solution for $\hGw$ for an arbitrary $\bU(z)$, but it transpires that the dispersion relation only depends on the part of the solution that \emph{can} be found. \Cref{eq:Greens_function_definition} has a particular solution in the form of vortices convected with the mean flow as shown by \citet{ellingsen2016waves}, but we do not need this for our purposes, and we consider the homogeneous solution $\hG_\mathrm{h}$ only. We consider the ranges above and one below the source separately and define
\begin{align} \label{eq:Greens_for_w}
    \hG_\mathrm{h}(z) = \left\{\begin{matrix}
        C_+\, \hGp (z)\quad & \text{ for } z>z_0\\
        C_-\, \hGm(z)\quad & \text{ for } z<z_0
    \end{matrix} \right. .
\end{align}
where $C_\pm$ are constants to be determined. The functions $\hG_\pm(z)$,  crucially, are both defined throughout the water column $z\in(-h,0)$ although at any given depth $\hG_\mathrm{h}$ is equal to one and not the other. They are smooth, and both satisfy the homogeneous Rayleigh equation everywhere. \Cref{eq:Bs_at_source} determines $C_\pm$ in terms of $\hG_\pm(z_0)$ and $\hG_\pm'(z_0)$ with solutions 
\begin{align}\label{eq:Gh}
    C_\pm  &= -\frac{1}{\cW(z_0) }\left[\hG_\mp'(z_0) + \frac{kU_\gamma'(z_0)}{\Omega(z_0)}\, \hG_\mp(z_0)\right].
\end{align}
The Wronskian is defined as $\cW(z;\bk,\omega)=\hGp'(z)\hGm(z)-\hGp(z)\hGm'(z) $. It is in fact a constant with respect to $z$. This can be seen when replacing $\hG''_\pm$ using the Rayleigh equation \eqref{eq:Greens_function_definition} for $z\neq z_0$. Therefore, we can set its value by evaluating it at the surface. We are free to choose a normalisation for $\hG_\mathrm{h}$, so we set
\begin{align} \label{eq:Greens_surface_values}
    \hGp(0) = 1,\qquad \Omz^2\, \hGp'(0) = \left[\big(g + \Upsilon k^2 \big)k^2 - \Omz k U_\gamma'(0) \right]\, \hGp(0) .
\end{align}
We use the boundary conditions in \eqref{eq:Greens_surface_values} to write
\begin{align}
\begin{split}
    \cW (z_0)=\cW (0) &= \hGp'(0)\, \hGm(0) - \hGp(0)\, \hGm'(0) 
    = - \frac{D(\bm{k}, \wint;\,\Ug)}{\wint^2},
\end{split}
\end{align}
where we used $\Omz=\wint$ and introduced the dispersion function
\begin{align} \label{eq:DISPERSION}
    D(\bm{k}, \wint;\,\Ug) \equiv \wint^2\, \hGm'(0) - \left[\big(g + \Upsilon k^2 \big) k^2 - k\wint\, U_\gamma'(0) \right]\, \hGm(0).
\end{align}
Noting that $D(\bk,\wint;\,\Ug)$ enters as a denominator in \eqref{eq:Gh}, we have the familiar structure which invariably appears in initial-value problems with surface waves, where the zeros of the denominator become poles in the inverse Fourier integrals that transform $\hw(\bk)\to w(\bx)$ etc, which by Cauchy's theorem pick out the $(\bk,\wint)$-combinations that allow freely propagating modes \citep[see, e.g.,][\S4.9]{Lighthill1978}. In other words, the dispersion relation is $D(\bk,\wint;\,\Ug)=0$. 

Note how only the bottom-normalised Green's function evaluated at the surface enters the dispersion relation. Because the boundary conditions at the bottom are of standard Dirichlet and Neumann type, a formal solution can be found in the region below the source and analytically continued to $z=0$.

%%%%%%%%%%%%%%%%% S E C T I O N %%%%%%%%%%%%%%%%%
\subsection{Solution below the source}

We next derive a form of $\hGm (z)$ that isolates the curvature $\Ug''(z)$ and is convenient for approximations and numerical implementation, using a procedure inspired by Dirac's `interaction picture' in quantum mechanics. The interaction picture is typically used for (quantum) systems evolving in time under a time-dependent forcing operator, provided the initial solution is known. Instead of time, we consider the `development' of $\hGm$ for $z>-h$ under the action of a $z$-dependent operator, when the solution at $z=-h$ is known. The idea is to split the operator acting on the unknown function into a sum of two parts, one containing the terms that give an exactly solvable solution, the second containing the `difficult' part, which is not, in general, solvable in closed form. 

The homogeneous Rayleigh equation that the Green's function $\hGm(z)$ adheres to reads 
\begin{equation}\label{eq:Rayleigh}
    \hGm ''(z)  = k^2\big[1 + q(z)\big]\, \hGm (z), \quad \text{ with } \quad \hGm(-h) = 0;\,\,\, \hGm'(-h) = \frac{k}{\cosh kh}.
\end{equation}
The factor $1/\cosh kh$ is an arbitrary choice, amounting to a choice of normalisation for $\hGm$ which ensures it tends sufficiently rapidly to zero as $kh\to\infty$, and the formalism transitions more smoothly across the full range of $kh$ values. Introducing the `state vector' ${\bcX(z) = \big(\hGm , \, \rk^{-1}\hGm ' \big)^\sT}$ ($\sT$: transpose) we re-cast \eqref{eq:Rayleigh} as
\begin{align} \label{eq:Rayleigh_G}
    \frac1\rk\bcX'(z)=
    \begin{pmatrix}
        0 & 1\\ 1 & 0
    \end{pmatrix}\bcX(z) 
    + \begin{pmatrix}
        0 & 0\\ q(z) & 0
    \end{pmatrix}\bcX(z) \equiv \bm{H}_0\bcX(z) + \bm{H}_U(z)\bcX(z),
\end{align}
defining the matrices $\bm{H}_0$ and $\bm{H}_U(z)$ as the first and second matrix in \eqref{eq:Rayleigh_G}, respectively. All effects of depth variation of $\Ug$ is now contained in the matrix $\bm{H}_U$. Inspired by quantum mechanics \citep[e.g.,][Ch.\ 6]{fetter2012quantum} we introduce a new vector $\bcY(z)$ defined by the linear transformation
\begin{subequations}
\begin{align}
    \bcY (z) &=\big(\cY_1, \, \cY_2 \big)^\sT \equiv \mT^{-1}(z+h)\, \bcX(z);\\ 
    \mT (x) &\equiv \rme^{\rk x\bm{H}_0 } =
    \begin{pmatrix}
        \cosh kx & \sinh kx\\ \sinh kx & \cosh kx
    \end{pmatrix}
    .\label{eq:defT}
\end{align}
\end{subequations}
The exponential is understood in a Taylor-series sense, $\rme^{\mA} = \bf{I}+\mA+\textstyle{\frac12}\mA^2+\dots$ with $\bf{I}$ the unit matrix, and the final form of \eqref{eq:defT} is obtained by evaluating the matrix exponential explicitly, e.g., equation 24 of \citep{blanes2009magnus}. Note that $\mT(-x)=\mT^{-1}(x)$ and that $\mT(0)=\bf{I}$, so $\bcY(z)\to\bcX(z)$ as $z\to-h$. The bottom boundary condition is $\bcX(-h)=\bcY(-h)=(0,1/\cosh kh)^\sT$. Inserting $\bcX=\mT\bcY$ into \eqref{eq:Rayleigh_G} the evolution equation becomes ${\bcY '(z) = k \bm{H}_Z(z)\,\bcY (z)}$, where 
\begin{align}\label{eq:Zevolution}
    \bm{H}_Z(z) \equiv \mT^{-1}(z+h)\,\bm{H}_U(z)\,\mT (z+h) = q(z)\,\mC(z+h),
\end{align}
with
\begin{align}
    \mC(x) = \frac{1}{2}
    \begin{pmatrix}
        -  \sinh 2k x & 1 - \cosh 2k x\\[3pt] 
          1 + \cosh 2k x & \sinh 2k x
    \end{pmatrix}.
\end{align}
The formal solution of $\bcY'= k \bm{H}_Z\bcY$ is then given by 
\begin{align} \label{eq:Y_state_integral_solution}
    \begin{pmatrix}
        \cY _1\\ \cY _2
    \end{pmatrix}
    =\cP\left\{\rme^{k\int_{-h}^z\rmd\zeta\,\bm{H}_Z(\zeta)}\right\}\bcY(-h)
        = \frac{1}{\cosh kh} \cP \left\{
        \rme^{\rk\!\int_{-h}^z \rmd\zeta\,q(\zeta)\,\mC(\zeta+h)}         
    \right\} \begin{pmatrix} 0\\1 \end{pmatrix} ,
\end{align}
where $\cP $ denotes the path ordering operator, which has to be applied since the matrix in the integrand does not commute when evaluated at two different depths, \citep[see, e.g.,][\S1]{dollard1979product}. We will discuss the interpretation of the ordering operator in \cref{sec:Dyson}, and in \cref{sec:Numeric} for a discretised domain and shear current, giving a natural approximation for numerical implementation. When there is a critical layer, i.e., when $\Ug(z_c) = \omega/k$ at some depth $z_c$, we shift the frequency off the real axis by a small amount $\omega \to \omega + \rmi\epsilon$ to represent a minuscule viscosity \citep{miles1957generation}, and take the limit $\epsilon \to 0^+$ after integration. 

The solution for $\hGm(z)$ and its derivative is given by ${\bcX(z) = \mT (z+h)\, \bcY(z)}$ which gives
\begin{align} \label{eq:Solution_bottom_Greens}
\begin{split}
        \hGm(z) &= \cosh k(z+h) \, \cY _1(z) + \sinh k(z+h) \cY _2(z), \\
        \hGm'(z) &= k \sinh k(z+h) \, \cY _1(z) + k \cosh k (z+h)\,  \cY _2(z) .
\end{split}
\end{align}
When inserting the Green's function evaluated at the surface into the dispersion \eqref{eq:DISPERSION}, the roots, representing freely propagating wave modes, are given by the equation
\begin{subequations}\label{eq:ZEROS_DISPERSION}
\begin{align} 
    \wint^2 =& \left[\big(g + \Upsilon k^2\big)\, k - \wint  \Ug'(0)\right] F(\bk, \wint) 
\end{align}
with 
\begin{align} \label{eq:Fraction_with_r}
    F(\bk, \wint) &= \frac{r + \tanh kh}{1 + r\tanh kh}; \\  
    r &= r(\bk,\wint)\equiv \frac{\cY_1(0)} {\cY_2(0)}.\label{eq:r}
\end{align}
\end{subequations}
\Cref{eq:ZEROS_DISPERSION} is the exact dispersion relation for arbitrary $U(z)$ from which $\wint(\bk)$ follows. As we will discuss, taking $r(\bk,\wint)=0$ reproduces the known relation for a linearly varying current, while the asymptotic behaviour of $r$ is analysed in Appendix \ref{sec:Asymptotic} and \ref{sec:DeepWater}. 

The fraction in \eqref{eq:Fraction_with_r} can be rewritten as $F=\tanh k(h + \tih)$, where $k\tih = \mathrm{artanh}\,r$. Therefore, a current with non-zero curvature gives $\tih\neq0$, which has the role of an effective depth change as a function of shear. This is an intuitive interpretation in shallow and intermediate water ($kh\lesssim O(1)$ and $|r|<1$).

%%%%%%%%%%%%%%%%% S E C T I O N %%%%%%%%%%%%%%%%%
\subsection{Deep-water limit}

The effective depth shift interpretation above is not applicable to deep water, and another form is needed. In the deep water limit, $kh \gg 1$, the hyperbolic functions behave as exponentials $\sinh kz, \cosh kz \sim e^{kz}$. Renormalising the path-ordered exponential yields a slightly modified solution (for details see Appendix \ref{sec:DeepWater}),
\begin{align}
\begin{split}
    \hGm(z) &= e^{kz} \big(\tilde{\cY}_1(z) + \tilde{\cY}_2(z) \big),\\
    \hGm'(z) &= - k e^{kz} \big(\tilde{\cY}_1(z) - \tilde{\cY}_2(z) \big),
\end{split}
\end{align}
where
\begin{align}
     \tbcY(z) = \mathcal{P} \exp \left\{ \frac{k}{2} \int_{-\infty}^z \rmd\zeta \begin{pmatrix}
        - 4 - q(\zeta) & - q(\zeta)\\ q(\zeta) & q(\zeta) 
    \end{pmatrix} \right\} \begin{pmatrix}
        0\\1
    \end{pmatrix}.
\end{align}
The deep water limit of \eqref{eq:ZEROS_DISPERSION} gives the same dispersion as found by \citet{shrira1993surface}, in which case 
\begin{equation}
    F(k, \wint)\buildrel{kh\to\infty}\over{\longrightarrow} \frac{1+R}{1-R},
\end{equation}
with
\begin{align} \label{eq:rTilde}
    R = \sum_{n=1}^\infty \left(-\frac{k}{2} \right)^n \int_{-\infty}^0 \!\!\!\rmd\zeta_1\, e^{k \zeta_1} q_1 \int_{-\infty}^0 \!\!\!\rmd \zeta_2\, e^{-k|\zeta_2 - \zeta_1|} q_2 \dots \int_{-\infty}^0 \!\!\!\rmd \zeta_n\, e^{-k|\zeta_n - \zeta_{n-1}|} q_n e^{k\zeta_n} ,
\end{align}
where we have used the shorthand $q_j \equiv q(\zeta_j)$.

%%%%%%%%%%%%%%%%% S E C T I O N %%%%%%%%%%%%%%%%%
\subsection{Dyson series representation} \label{sec:Dyson}

A potentially instructive form of \eqref{eq:Y_state_integral_solution} is obtained by casting the path-ordered integral in another form. We may expand the exponential in orders of $\bm{H}_Z(z) =  q(z) \bm{C}(z+h)$, keeping the ordering 
\begin{align} \label{eq:Dyson}
    &\bcY(z) = \frac{1}{\cosh kh} \left[\bm{I} +\sum_{n=1}^\infty \frac{k^n}{n!} \int_{-h}^z \!\rmd\zeta_1 \cdots\int_{-h}^z \!\rmd\zeta_n \mathcal{P}\{\bm{H}_Z(\zeta_1) \cdots \bm{H}_Z(\zeta_n) \} \right] \begin{pmatrix} 0\\1 \end{pmatrix}.
\end{align}
The factor $1/n!$ assures convergence for all finite $kh$. At each order $n$, the solution \\ \noindent ${\bcX^{(n)}(z) = \mT(z+h) \bcY^{(n)}(z)}$ is proportional to $e^{kh}$, asymptotically for $kh\gg1$, which is balanced by the factor $1/\cosh kh$ in the boundary condition $\bcY(-h)$. We show this in Appendix \ref{sec:DysonAsymptote}. The path-ordering operator places the largest coordinate $\zeta_j$ to the left, $\mathcal{P} \bm{H}_Z(\zeta_1) \bm{H}_Z(\zeta_2) = \bm{H}_Z(\zeta_i) \bm{H}_Z(\zeta_j)$ with $i,j\in\{1,2\}$, such that $\zeta_i>\zeta_j$. An alternative way to write \eqref{eq:Dyson} makes the path ordering explicit, 
\begin{align} \label{eq:Dyson2}
    \bcY(z) = \frac{1}{\cosh kh}\left[\bm{I} + \sum_{n=1}^{\infty} k^n \int_{-h}^z \!\rmd \zeta_1 \int_{-h}^{\zeta_1} \!\rmd \zeta_2 \cdots \int_{-h}^{\zeta_{n-1}} \!\!\rmd \zeta_n\, \prod_{j=1}^n \bm{H}_Z(\zeta_j) \right] \begin{pmatrix} 0\\1 \end{pmatrix}.
\end{align}
Equations \eqref{eq:Dyson} and \eqref{eq:Dyson2} do not require weak curvature, but readily yield a weak-curvature approximation when truncated, as we discuss in section \ref{sec:weakApprox}. It is useful for numerical evaluation of \eqref{eq:ZEROS_DISPERSION} since convergence is usually rapid in practical problems.

Curiously, equations \eqref{eq:Dyson} and \eqref{eq:Dyson2} have the form of a Dyson series \citep{Dyson1949}, well known from quantum physics to describe many-particle interactions via multiple scattering, providing powerful Green's function solutions for a wide range of problems \citep[e.g.][]{Economou2006} able to tackle highly complex geometries \citep[e.g.][]{Ellingsen2011} and even describe the scattering of surface waves on turbulent eddies \citep{Sazontov1982}, allowing the intuitive use of Feynman diagrams. Each term then represents, in analogy, increasing number of internal scatterings of curvature interacting with itself at different depths via the bottom; the second-order term, say, has integrand of form $q(\zeta_1)\mC(\zeta_1+h)\mC(\zeta_2+h)q(\zeta_2)$, with $\mC$ akin to a propagator of a `signal' from $\zeta_1$ to $-h$ and thence to $\zeta_2$. Higher-order terms describe repeated scatterings, whereby curvature at one depth first generates a correction that is then acted upon again by curvature at shallower depths, as seen in the nested integrals. Note also that self-interactions (i.e.\ $\zeta_1=\zeta_2$) are prohibited by the nilpotency of the integrand, $\mC(x) \mC(x) = 0$; one readily verifies that $\mC(\zeta_1+h)\mC(\zeta_2+h)\propto \sinh k(\zeta_1-\zeta_2)$. 

We leave the `multiple scattering analogy' as a curious observation at present.

%%%%%%%%%%%%%%%%%%%%%%%%%%%%%%%%%%%%%%%%%%%%%%%%%
%%%%%%%%%%%%%%%%% S E C T I O N %%%%%%%%%%%%%%%%%
%%%%%%%%%%%%%%%%%%%%%%%%%%%%%%%%%%%%%%%%%%%%%%%%%
\section{Known analytical solutions and approximations} \label{sec:Classical_and_approximations}

In this section, we show that equation \eqref{eq:ZEROS_DISPERSION} reduces to known exact dispersion relations and yields well-known approximate formulae.

%%%%%%%%%%%%%%%%% S E C T I O N %%%%%%%%%%%%%%%%%
\subsection{Exact analytical limits}
The most classical case of a depth-varying current is that with constant shear, i.e., $\bU(z) = (U_0+Sz)\bde_U$, with constant $U_0$ and vorticity $S$. In this case we have $q(z)=0$ which leads to $\bm{H}_Z=0$, and in turn $\bcY = (0,1/\cosh kh)^\sT$ and $r(\bk,\omega)=0$ in \eqref{eq:ZEROS_DISPERSION}, which, when defining $\bk=k[\cos\theta,\sin\theta]$, yields the classical dispersion relation \citep{craik1968}
\begin{align}\label{eq:constShear}
    \wint^2 + \wint  S \cos\gamma \tanh kh -\omega_0^2  =0,
\end{align} 
with $\cos\gamma=\bk\cdot\bde_U/k$ as before and $\wint=\omega-kU_0\cos\gamma$. 

Secondly, we consider a class where ${\Ug''(z) = a^2 \Ug(z)}$, studied in detail by \citet[][\S IV.B.2]{peregrine1976interaction}. Since $a$ can be complex, this form encompasses a range of exponentially and trigonometrically depth dependent profiles. They have a simple analytical solution for one wavenumber value $k=k_0$ which satisfies ${\omega(k_0) = 0}$, representing a stationary wave or a `travelling wave' of constant form if its phase speed is assumed known. When defining an effective wavenumber $\kappa^2 = k_0^2 + a^2$, the Rayleigh equation \eqref{eq:Rayleigh_w} (homogeneous form) is $w''(z) - \kappa^2 w (z) = 0$, immediately resulting in \eqref{eq:Rayleigh_G} with $k$ replaced by $\kappa$ and $q=0$. An equation fixing the value of $k_0$ must then be found. Since $\kappa$ is not depth-dependent, we can insert it into \eqref{eq:Solution_bottom_Greens}, which gives the ordinary solution ${\hGm (z) =k \sinh \kappa(z+h) /\kappa \cosh kh}$. In this case the curvature measure $q$ of equation \eqref{eq:q} reduces to $q=a^2/k^2$, and inserting this into the dispersion relation, we find the equation implicitly defining $k_0$,
\begin{align} \label{eq:Peregrine_wavenumber}
    \Ug(0)^2\,\kappa \coth\kappa h = g + \Upsilon k_0^2 + \Ug(0)\,\Ug'(0),
\end{align}
identical to equation (4.21) of \citet{peregrine1976interaction} (but for our inclusion of surface tension).

%%%%%%%%%%%%%%%%% S E C T I O N %%%%%%%%%%%%%%%%%
\subsection{Approximations for weak shear and weak curvature} \label{sec:weakApprox}

As mentioned in the Introduction, the most commonly used dispersion relations for practical applications are explicit, analytical approximations. The approximations of \cite{Stewart1974} (infinite depth), \cite{skop1987approximate} (finite depth), and \cite{kirby1989surface} (second order) all require that the shear is weak. Specifically, it was shown by \cite{ellingsen2017approximate} that they require $|s(\bk)|\ll 1$ where the wave--shear-strength number $s$ was defined in equation \eqref{eq:s}. Another form was derived by \citet{ellingsen2017approximate}, allowing $s$ to be arbitrarily large, provided the depth-averaged curvature $\Ug''$ is small in the appropriate sense. In that case, $\wint\approx \omega_\mathrm{w.crv}$ with
\begin{equation}\label{eq:ELdisp}
 \omega_\mathrm{w.crv}\equiv (\sqrt{1+s^2}-s)\omega_0. 
\end{equation}
The weak-shear approximation \citep{skop1987approximate} follows from \eqref{eq:ELdisp} by retaining the leading order term in $s$, which is seen when noting that this approximation may be written as $\wint\approx \omega_\mathrm{w.sh}$ with 
\begin{equation}\label{eq:weakshear}
    \omega_\mathrm{w.sh}\equiv (1-s)\omega_0 = \omega_0 - \int_{-h}^0\rmd z\,\frac{k\sinh 2k(z+h)}{\sinh2kh}\Ug'(z),
\end{equation}
\citep[see][]{ellingsen2017approximate}, which is found by truncating to first order in $s$.  A better known equivalent form is related to \eqref{eq:weakshear} via a partial integration, 
\begin{equation}\label{eq:weakshear_doppler}
    \omega(\bk) = \omega_0(k) + \int_{-h}^0\rmd z\,\frac{2k^2 \cosh 2k(z+h)}{\sinh2kh}\Ug(z),
\end{equation}
often understood as an ``effective Doppler-shift" as in equation \eqref{eq:effDopp} \citep[the ``Doppler shift" understanding has unfortunately led to some misunderstandings with regards to group velocity when implemented in ocean models as explained by][]{banihashemi2017}.

We can obtain the weak-curvature approximation from our dispersion in \eqref{eq:ZEROS_DISPERSION}; We begin by multiplying the Rayleigh equation in \eqref{eq:Rayleigh} by $\sinh k(z+h)$ and integrating over the vertical domain. This gives us a relation between $\hGm'(0)$ and $\hGm (0)$. In the spirit of \citet{shrira1993surface} we make a ‘near-potentiality' assumption that the path-ordered exponential in \eqref{eq:Y_state_integral_solution} may be expanded in powers of $q(z)$, meaning that the effect of curvature is sufficiently small to ensure convergence of the Dyson series \eqref{eq:Dyson}. Truncating the series after $n=1$ terms as detailed in Appendix \ref{app:app_DIM}, yields the weak-curvature-relation found by \citet{ellingsen2017approximate},
\begin{align} \label{eq:EL_dispersion}
    \wint^2 + 2\, \wint\,\omega_0\, s- \omega_0^2 + \omega_0^2\, \Delta = 0,
\end{align}
where we have collected the curvature effects into a dimensionless quantity
\begin{align} \label{eq:Delta}
    \Delta(\bk) \equiv \frac{4 \sigma k^3}{\omega_0^2}\!\! \int_{-h}^0 \!\!\rmd z\,  q(z) \frac{\sinh k(z+h)}{ \sinh 2kh} \int_z^0 \!\!\rmd \zeta\, \big[\langle\Ug\rangle-\Ug(\zeta)\big] \cosh k(2 \zeta +h -z).
\end{align}
When the curvature is weak, $\Delta\ll s$, we get $\wint\to\omega_\mathrm{w.crv}$ which tends to $\omega_\mathrm{w.sh}$ when $s\ll 1$. Thus, all these approximations are contained in the dispersion relation \eqref{eq:ZEROS_DISPERSION}.

%%%%%%%%%%%%%%%%%%%%%%%%%%%%%%%%%%%%%%%%%%%%%%%%%
%%%%%%%%%%%%%%%%% S E C T I O N %%%%%%%%%%%%%%%%%
%%%%%%%%%%%%%%%%%%%%%%%%%%%%%%%%%%%%%%%%%%%%%%%%%
\section{Numerical implementation} \label{sec:Numeric}

In this section, we numerically evaluate $\sigma(\bk)$ using our dispersion relation to validate it in cases with nonzero curvature, which pose additional mathematical challenges. In particular, the solution in \eqref{eq:Y_state_integral_solution} involves a path-ordered exponential, a generalisation of the ordinary exponential solution. The key difficulty arises because, for non-commuting matrices, the order in which terms are multiplied along the integration path affects the result, preventing a simple closed-form solution. There are many ways to evaluate our expression numerically, including Magnus expansions \citep{blanes2009magnus} and ordered-product schemes \citep{dollard1979product}; we focus here on the latter, since it is easy to implement. 

To clarify the problem at hand, consider the following scalar problem, ${F'(z) = A(z)\, F(z)}$, where the solution is readily given by $F(z) = F(-h)\exp[\int_{-h}^z \rmd \zeta\, A(\zeta)]$. In contrast, if the problem involves matrices, $\bcF'(z)=\mA(z) \bcF(z)$ where the elements of $\mA$ have different $z$-dependence, the solution is not simply the matrix exponential $\exp[\int_{-h}^z \rmd \zeta\, \mA(\zeta)]$, but one where the Taylor expansion is replaced by a series of nested integrals of $\mA$ involving matrix multiplications of $\mA$ with itself at different depths which do not in general commute and must be performed in the right order. The path-ordering operator ensures that the sequence of multiplications correctly reflects the underlying mathematics. 

Alternatively to the Dyson series \eqref{eq:Dyson}, the path-ordered exponential can be interpreted as the continuum limit of a discrete product of propagators—each corresponding to a thin layer of thickness $\Delta z = h/N$
of the $z$ domain. In this approach, the solution is constructed as an ordered product,  
\begin{align} \label{eq:Disc_limit}
\begin{split}
        \bcY(0) &= \lim_{N\to\infty} \left( \prod_{n^*=0}^{N-1} e^{k\, \bm{H}_Z(z_n)\, \Delta z} \right) \bcY(-h) \\
        &= \frac{1}{\cosh kh} \lim_{N\to\infty} \prod_{n^*=0}^{N-1} \bigg(\bm{I} + k\, \bm{H}_Z(z_n)\, \Delta z \bigg)\begin{pmatrix}0\\1\end{pmatrix},
\end{split}
\end{align}
where $z_n = n\Delta z - h$. Here, the product operator acts in reverse to preserve the correct multiplication order: $\prod_{n^*} A_n = A_{N-1} \dots A_1 A_0$. This discrete viewpoint underpins our numerical implementation: we use \eqref{eq:Disc_limit} with $N$ high enough to reach the desired tolerance and approximate $\bm{H}_Z(z_n)$ at the midpoint $\tfrac{1}{2}(z_n + z_{n+1})$. The last equality in \eqref{eq:Disc_limit} is obtained since the exponent in the infinitesimal propagator is nilpotent, $\mC^2=0$. Finally, we employ a bisection method to solve for the root of $D(k,\omega)$ with $M$ refinements.

\begin{figure}
    \centering
    \includegraphics[width=\textwidth]{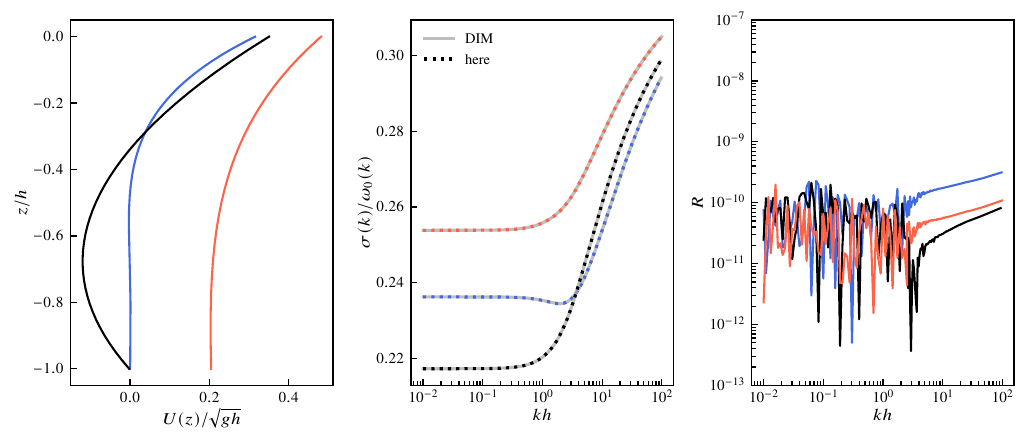}
    \caption{Three vertical shear profiles and their dispersion relations found with DIM and the expression derived herein, equation \eqref{eq:ZEROS_DISPERSION}. In the third column, we show the relative error $R$ between our dispersion and DIM.}
    \label{fig:Verification_DIM}
\end{figure}

We compare our implementation of the dispersion relation given in \eqref{eq:ZEROS_DISPERSION} with a numerical solver using the Direct Integration Method (DIM) \citep{li2019framework}, providing comparison with an extensively validated method. The velocity profiles used for this comparison are sixth-order polynomial curves designed to closely mimic the complex structure of wind-induced surface currents, as used by \citet{swan2000simple}. These profiles capture the gradual variation and curvature observed in realistic geophysical flows. We used $N=10^5$ so that the layerwise approximation of the path-ordered exponential was sufficiently fine for the computed roots to be insensitive to further refinement of the vertical grid. For both methods, we used $M=50$ bisection refinements to robustly compute the roots of the dispersion relation.

The dispersion curves shown in \cref{fig:Verification_DIM} are computed for wavevectors parallel to the underlying shear current ($\gamma=0$) with no loss of generality. The first column of the figure visually presents the shear velocity profiles, highlighting the pronounced curvature and depth dependence captured by the sixth-order polynomials. Next to these, the corresponding intrinsic frequency dispersion relations $\wint(k)/\omega_0(k)$ are plotted, illustrating how each velocity profile modifies the wave characteristics. In the final column, we quantitatively assess the numerical performance by plotting the relative error ${R = |\wint - \wint^{\textrm{DIM}}| / \wint^{\textrm{DIM}}}$ between our results and those obtained by DIM. Across all cases, our dispersion relation and DIM give identical results, limited only by numerical precision. These test cases serve as benchmarks, and we have extensively tested a wide variety of velocity profiles, with full agreement in all cases. However, it is important to emphasise that the primary purpose of these comparisons is not to verify our numerical implementation, but to demonstrate the correctness of the general dispersion relation we have derived. A computer code demonstrating implementation of the method for the profiles of Fig. \ref{fig:Verification_DIM} is included as Supplementary Material.

%%%%%%%%%%%%%%%%%%%%%%%%%%%%%%%%%%%%%%%%%%%%%%%%%
%%%%%%%%%%%%%%%%% S E C T I O N %%%%%%%%%%%%%%%%%
%%%%%%%%%%%%%%%%%%%%%%%%%%%%%%%%%%%%%%%%%%%%%%%%%
\section{Conclusions}

We have found a closed, implicit dispersion relation for linear surface waves on an arbitrary shear current. Unlike classical treatments that expand in global eigenmodes satisfying both the bottom and free-surface boundary conditions, we instead constructed the Green's function from two fundamental solutions: one satisfying the bottom condition and the other the free-surface condition. As a result, the dispersion relation follows from the Wronskian, which involves only the bottom solution evaluated at the surface. This approach eliminates the eigenfunction from the formulation, allowing the exact dispersion to be written as a closed functional of the prescribed shear profile. The dispersion relation involves nested, path-ordered integrals and is equivalent to that by \citet{shrira1993surface} in the deep-water limit. Curiously it may be expressed as a Dyson series expansion akin to multiple scattering in quantum physics with the curvature $U''$ at different depths acting as ``particles''.

To validate our approach, we showed that the dispersion relation reproduces known analytical solutions for constant vorticity and a class of exact solutions due to \citet{peregrine1976interaction}, and connects to known approximations for weak shear and weak curvature. Additionally, we numerically implemented the dispersion relation and systematically tested it across a broad spectrum of velocity profiles, ranging from simple monotonic shears to flows with multiple inflection points and currents that mimic realistic scenarios. In every case, the computed dispersion curves were identical to those obtained using DIM \citep{li2019framework}, limited only by numerical precision. This comprehensive agreement across diverse and challenging profiles underscores the robustness of our analytical approach and its applicability to a wide range of geophysical flow scenarios. Taken together, these findings demonstrate that the analysis presented here provides a fully self-contained one-equation formal solution to the dispersion relation $\omega=\omega(k)$ for linear surface waves on horizontally directed currents $\bU(z)=[U(z),V(z),0]$ with arbitrary vertical structure. The new solution can serve as a starting point for future analytical work, or provide an exact benchmark solution.

\section*{Funding.}
The research was funded by the European Union (ERC CoG, \emph{WaTurSheD}, grant 101045299). Views and opinions expressed are, however, those of the authors only and do not necessarily reflect those of the European Union or the European Research Council. Neither the European Union nor the granting authority can be held responsible for them.

\section*{Acknowledgements.}
KSH thanks H. H. Fyhn for an interesting discussion that inspired this investigation.

\appendix
%%%%%%%%%%%%%%%%%%%%%%%%%%%%%%%%%%%%%%%%%%%%%%%%%%%
%%%%%%%%%%%%%%%%% A P P E N D I X %%%%%%%%%%%%%%%%%
%%%%%%%%%%%%%%%%%%%%%%%%%%%%%%%%%%%%%%%%%%%%%%%%%%%
\appendix

%%%%%%%%%%%%%%%%% S E C T I O N %%%%%%%%%%%%%%%%%
\section{Shallow-water limit} \label{sec:Asymptotic}

To find appropriate limits of the dispersion relation in \eqref{eq:ZEROS_DISPERSION}, we target the fraction $r$. Instead of expanding the path-ordered exponential, we analyse the Riccati equation for $r(z)$, given by
\begin{align} \label{eq:Ricatti_for_r}
    r'(z) = - k\, q(z)\, \big[r(z) \cosh k\xi + \sinh k\xi \big]^2\,.
\end{align}
with shorthand $\xi=z+h$. This equation can be found from the equation ${\bcY '(z) = k \mH_Z(z)\,\bcY (z)}$. The original fraction $r$ is then the value at $z=0$. 

We consider the shallow-water asymptote $\vk\equiv kh\ll1$, implying $k\xi=\vk\bsig\ll1$ where we introduce $\bsig=\xi/h$, so $0\leq\bsig\leq 1$.  Equation \eqref{eq:Ricatti_for_r} implies
\begin{align}\label{eq:riccr}
    r'(z) \approx \frac{\Ug''(z)}{\omega - k\Ug(z)}\big[ r(z) + \vk\bsig + \tfrac{1}{2}\vk^2\bsig^2 r(z)\big]^2 + \order{\vk^3}.
\end{align}
Dominant balance shows that $r\sim \order{\vk}$ or smaller \citep[see, e.g.][\S 3.4]{Bender1999}, so we write $r(z)=\vk \vrho(\bsig)$, define $\cU(\bsig)=k\Ug(z)/\omega_0$ and write \eqref{eq:riccr} in terms of $\bsig$, 
\begin{equation}
    \vrho'(\bsig)=\frac{\cU''(\bsig)}{\omega/\omega_0-\cU(\bsig)}[\vrho(\bsig)+\bsig]^2 +\order{\vk}.
\end{equation}
Integrating the equation for $\vrho(\bsig)$ gives (noting that $\vrho(0)=0$)
\begin{align}
    \vrho(\bsig) = \int_0^{\bsig}Q(\tau)\,[\vrho(\tau)+\tau]^2  \rmd\tau, \qquad Q(\tau)\equiv \frac{\cU''(\tau)}{\omega/\omega_0-\cU(\tau)}.
\end{align}
Now we write the solution as a series in powers of $\lambda$ (which we set to 1 in the end), and multiply $Q$ by it as well, so the solution becomes iterative. We write the solution as a Volterra-type series
\begin{align}
    \vrho(\bsig) = \sum_{n=1} \lambda^n \vrho^{(n)}(\bsig),
\end{align}
where the different powers are defined by
\begin{align}
\begin{split}
        \sum_{n=1} \lambda^n \vrho^{(n)}(\bsig) =& \int_0^{\bsig} \rmd \tau\, Q(\tau) \bigg[\lambda \tau^2 \\
        &+ 2 \tau \sum_{n=1} \lambda^{n+1} \vrho^{(n)}(\tau) + \sum_{m=2} \lambda^{m+1} \sum_{k=1}^{m-1} \vrho^{(k)}(\tau)\, \vrho^{(m-k)}(\tau)\bigg].
\end{split}
\end{align}
So, in summary, the formal solution is $r = \vk\vrho(1)$ where
\begin{align}
\begin{split}
    \vrho(\bsig)= \sum_{n=1}^\infty \vrho^{(n)}(\bsig),&\qquad \vrho^{(1)}(\bsig) = \int_0^{\bsig} \rmd\tau\, \tau^2 Q(\tau), \\
    \vrho^{(n+1)}(\bsig) =  2 \int_0^{\bsig} \rmd\tau\, \tau\, Q(\tau)\, \vrho^{(n)}&(\tau) + \sum_{k=1}^{n-1} \int_0^{\bsig} \rmd\tau\, Q(\tau)\, \vrho^{(k)}(\tau)\, \vrho^{(n-k)}(\tau) .
\end{split}
\end{align}

%%%%%%%%%%%%%%%%% S E C T I O N %%%%%%%%%%%%%%%%%
\section{Deep-water limit} \label{sec:DeepWater}

To obtain the deep-water limit, we return to \eqref{eq:Rayleigh_G} and introduce $\mTinf(z) \equiv \mT(z) \mV$, where $\mV$ is a constant invertible matrix chosen to diagonalise $\mH_0$. Since $\mT'(z) = k \mH_0 \mT(z)$ and $\mH_0 \mV  = \mV\boldsymbol{\Lambda}$, we have $\mTinf'(z) =  k \mTinf(z) \boldsymbol{\Lambda}$. Choosing $\mV$ from the eigenvectors of $\mH_0$, we write
\begin{align}
    \mV = \begin{pmatrix}
        -1&1\\1&1
    \end{pmatrix},\quad \boldsymbol{\Lambda} = \begin{pmatrix}
        -1&0\\0&1
    \end{pmatrix}.
\end{align}
With this choice, the new operator becomes
\begin{align}
    \mTinf(z) = \begin{pmatrix}
        -e^{-kz}&e^{kz}\\e^{-kz}&e^{kz}
    \end{pmatrix}, \quad \mCinf(z) = \frac{1}{2} \begin{pmatrix}
        -1&e^{2 k z}\\- e^{-2 k z}&1
    \end{pmatrix},
\end{align}
where $\mCinf$ is defined by $q(z)\mCinf(z)\equiv\mTinf^{-1}(z) \mH_U(z) \mTinf(z)$, as before. We define the state $\bcY_\infty(z) \equiv \mTinf^{-1}(z) \bcX(z)$, which satisfies the equation $\bcY_\infty'(z) = k q(z) \mCinf(z) \bcY_\infty(z)$. The next step is to renormalise $\bcY_\infty(z)$. Since only the ratio of the two components enters the dispersion relation through $r$ in \eqref{eq:ZEROS_DISPERSION}, any common factor is irrelevant. We therefore set 
\begin{align}
    \tbcY(z) \equiv \mR(z) \bcY_\infty(z), \quad \mR(z) = \begin{pmatrix}
        -e^{-2kz} & 0 \\ 0 & 1
    \end{pmatrix}.
\end{align}
Substituting this into the evolution equation yields 
\begin{align}
    \tbcY'(z) = \frac{k}{2} \begin{pmatrix}
        - 4 - q(z) & - q(z)\\ q(z) & q(z) 
    \end{pmatrix} \tbcY(z) .
\end{align}
The boundary condition becomes $\tbcY(-\infty) = (0,\, 1)^\sT$, so the formal solution can be written as
\begin{align}
    \tbcY(z) = \mathcal{P} \exp \left\{ \frac{k}{2} \int_{-\infty}^z \rmd\zeta \begin{pmatrix}
        - 4 - q(\zeta) & - q(\zeta)\\ q(\zeta) & q(\zeta) 
    \end{pmatrix} \right\} \begin{pmatrix}
        0\\1
    \end{pmatrix}.
\end{align}
To recover the Green's function, we use $\bcX(z) = \mTinf(z) \bcY_\infty(z)$ together with $\cY_{\infty,1} = -e^{2 k z} \tilde{\cY}_1$ and $\cY_{\infty,2} = \tilde{\cY}_2$. Substituting into the expressions for $\hGm$ and $\hGm'$ gives
\begin{align}
    \hGm(z) = e^{kz} \big(\tilde{\cY}_1(z) + \tilde{\cY}_2(z) \big),\qquad
    \hGm'(z) = - k e^{kz} \big(\tilde{\cY}_1(z) - \tilde{\cY}_2(z) \big).
\end{align}
Finally, defining $R = \tilde{\cY}_1(0)/\tilde{\cY}_2(0)$, and ignoring common factors, we obtain $\hGm = (1+R)$ and $ \hGm' = k (1 - R)$, which is the same deep-water structure found by \citet{shrira1993surface}. Note that $R$ is \textit{not} the deep-water limit of $r$.

%%%%%%%%%%%%%%%%% S E C T I O N %%%%%%%%%%%%%%%%%
\section{Dyson series asymptote} \label{sec:DysonAsymptote}
The full solution for the Green's function $\hGm(z)$ and its derivative is given by
\begin{align}\label{eq:DysonX}
    \bcX(z) = \mT(z+h) \frac{1}{\cosh kh} \left[\bm{I} + \sum_{n=1}^{\infty} k^n \int_{-h}^z\!\rmd\zeta_1\cdots\int_{-h}^{\zeta_{n-1}} \!\rmd\zeta_n \prod_{j=1}^n \mH_Z(\zeta_j)
    \right] \begin{pmatrix}0\\1\end{pmatrix},
\end{align}
where 
\begin{align}
    \prod_{j=1}^n \mH_Z(\zeta_j) = \left(\prod_{j=1}^n  q(\zeta_j)\right) \left( \prod_{j=1}^n \mC(\zeta_j+h)\right).
\end{align}
We note that $\mC(\zeta+h)\propto e^{2kh}$, so that the $n$th term falsely appears $\propto e^{2nkh}$ which would cause the series to diverge for $kh\gg1$. However, it is readily shown from equation \eqref{eq:Zevolution} that one can rewrite $\mC(\zeta+h) = \mT(-h) \mC(\zeta) \mT(h)$, where $\mT(h)\mT(-h) = \bm{I}$ since $\mT(a+b) = \mT(a)\mT(b)$ and $\mT(0)=\bm{I}$ from equation \eqref{eq:defT}. Therefore, the integrand in equation \eqref{eq:DysonX} is proportional to
\begin{align}
    \mT(z+h) \prod_{j=1}^n \mC(\zeta_j+h) 
    =  \mT(z+h) \prod_{j=1}^n\left[ \mT(-h) \mC(\zeta_j)\mT(h)  \right] 
    = \mT(z) \left(\prod_{j=1}^n \mC(\zeta_j) \right) \mT(h),
\end{align}
which is $\propto e^{kh}$, and the normalising prefactor $1/\cosh kh$ ensures that $\bcX$ does not become exponentially large in the deep-water limit. The product of $\mC$ matrices gives
\begin{align} \label{eq:prod_C}
    \prod_{j=1}^n \mC(\zeta_j) = \left(\prod_{j=1}^{n-1}  \sinh k(\zeta_j - \zeta_{j+1})\right) \begin{pmatrix}
        -\sinh k\zeta_1\, \cosh k\zeta_n & -\sinh k\zeta_1\, \sinh k\zeta_n \\ \cosh k\zeta_1\, \cosh k\zeta_n & \cosh k\zeta_1\, \sinh k\zeta_n
    \end{pmatrix}.
\end{align}
Multiplying \eqref{eq:prod_C} from the left by $\mT(z)$ and from the right by $\mT(h) (0,\,1)^\sT$, we obtain
\begin{align}
\begin{split}
        \mT(z) &\prod_{j=1}^n \mC(\zeta_j) \mT(h) \begin{pmatrix} 0\\1\end{pmatrix} \\
    &=  \left(\prod_{j=1}^{n-1}  \sinh k(\zeta_j - \zeta_{j+1})\right) \sinh k(\zeta_n + h) \begin{pmatrix}
        \sinh k(z-\zeta_1) \\ \cosh k(z-\zeta_1)
    \end{pmatrix},
\end{split}
\end{align}
This expression shows explicitly how the apparent growth cancels. Expanding the hyperbolic functions in exponentials, we may rewrite the integrand as
\begin{align}
        &\sinh k(\zeta_n+h) \left(\prod_{j=1}^{n-1} \sinh k(\zeta_j - \zeta_{j+1}) \right) \begin{pmatrix}
        \sinh k(z-\zeta_1) \\ \cosh k(z-\zeta_1)
    \end{pmatrix} \\
        &= \frac{e^{k (\zeta_n+h)}}{2} \bigg(1 - e^{-2k(\zeta_n+h)}\bigg) \left(\prod_{j=1}^{n-1} \frac{e^{k(\zeta_j - \zeta_{j+1})}}{2} \bigg(1-e^{-2k(\zeta_j - \zeta_{j+1})}\bigg) \right)  \frac{e^{k (z-\zeta_1)}}{2} \begin{pmatrix}
        1 - e^{-2k(z-\zeta_1)} \\ 1 + e^{-2k(z-\zeta_1)}
    \end{pmatrix} . \nonumber
\end{align}
Since the ordered integrations impose $-h\leq\zeta_n\leq\zeta_{n-1}\leq\cdots\leq\zeta_1\leq z$, all arguments of the factors $1-e^{-2k(\cdot)}$ are non-negative. Therefore, these factors are bounded $0\leq1-e^{-2k(\cdot)}\leq1$, while the exponentials become
\begin{align}
    e^{k(z-\zeta_1)}e^{k(\zeta_1 - \zeta_2)}e^{k(\zeta_2 -\zeta_3)} \cdots e^{k(\zeta_{n-1} - \zeta_n)} e^{k(\zeta_n +h)} = e^{k(z+h)} .
\end{align}
Therefore, the integrand in the $n$'th term becomes 
\begin{align}
\begin{split}
        &\frac{k^n}{\cosh kh} \mT(z) \prod_{j=1}^n q(\zeta_j)\mC(\zeta_j) \mT(h) \begin{pmatrix} 0\\1\end{pmatrix} \\
        &= \frac{k^n e^{kz}}{2^n} \frac{1 - e^{-2k(\zeta_n+h)}}{1 + e^{-2 k h}} \prod_{j=1}^{n-1} \left(1-e^{-2 k(\zeta_j - \zeta_{j+1})} \right)  \prod_{\ell=1}^n q(\zeta_\ell)\begin{pmatrix}
        1 - e^{-2k(z-\zeta_1)} \\ 1 + e^{-2k(z-\zeta_1)}
    \end{pmatrix}.
\end{split}
\end{align}

%%%%%%%%%%%%%%%%% S E C T I O N %%%%%%%%%%%%%%%%%
\section{Weak curvature approximation}\label{app:app_DIM}

Integrating the left-hand side of the Rayleigh equation \eqref{eq:Rayleigh} multiplied by sinh by parts twice yields
\begin{align}
\begin{split}
         \int_{-h}^0 \mathrm{d}z\, \hGm''(z)\, \sinh k(z+h) = &\hGm'(0) \sinh kh -\hGm(0) k \cosh kh \\
     & +\int_{-h}^0 \mathrm{d}z\, \hGm(z)\, k^2 \sinh k(z+h)
\end{split}
\end{align}
where we used that $\hGm(-h) = 0$.
The $k^2$ term cancels on each side of the integrated Rayleigh equation. Thus we can express the derivative of the Green's function at the surface as
\begin{align}
    \hGm'(0) = \hGm(0) k \coth kh + \int_{-h}^0 \mathrm{d}z\, \hGm(z)\, k^2\,q(z)\frac{\sinh k(z+h)}{\sinh kh},
\end{align}
which, when inserted into $D(\bm{k}, \wint;\Ug)$ in equation \eqref{eq:DISPERSION} gives dispersion relation
\begin{align} \label{eq:Dispersion_before_weak_curvature}
        &\wint^2 - \omega_0^2 + \wint \Ug'(0) {\tanh kh} = \int_{-h}^0 \mathrm{d}z\, \Ug''(z) \left[\wint + \wint\frac{k\Delta\Ug(z)}{\Omega(z)} \right] \frac{\hGm(z)}{\hGm(0)} \frac{\sinh k (z+h)}{\cosh kh},
\end{align}
where $\Delta\Ug(z)\equiv\Ug(z)-\Ug(0)$. Next, we rewrite the shear as a boundary term obtained by integrating by parts,
\begin{align}
\begin{split}
        \Ug'(0) \tanh kh =&\int_{-h}^0 \rmd z\, \Ug''(z) \frac{\hGm(z)}{\hGm(0)} \frac{\sinh k(z+h)}{\cosh kh} \, \\
        &+ \int_{-h}^0 \rmd z\, \Ug'(z)\,\frac{d}{dz} \left(\frac{\hGm(z)}{\hGm(0)} \frac{\sinh k(z+h)}{\cosh kh} \right)  .
\end{split}
\end{align} 
The weak curvature approximation is obtained by truncating the Dyson series for $\bcY(z)$ in \eqref{eq:Dyson} at the first order,
\begin{align}
    \bcY(z) \approx {\frac{1}{\cosh kh}}\left\{\bm{I} + k \int_{-h}^z \rmd \zeta \, q(\zeta) \mC(\zeta+h) \right\} \begin{pmatrix}0\\1\end{pmatrix},  
\end{align}
so that
\begin{align}
    \begin{split}
        \cY_1(z) &\approx -k \int_{-h}^z \rmd \zeta\, q(\zeta) \frac{\sinh^2k(\zeta+h)}{{\cosh kh}};\\
        \cY_2(z) &\approx \frac{1}{\cosh kh} + \frac{k}{2} \int_{-h}^z \rmd \zeta \, q(\zeta) \frac{\sinh 2k (\zeta+h)}{{\cosh kh}},
    \end{split}
\end{align}
when noting that $1-\cosh 2kx=-2\sinh^2 kx$. 
Therefore we insert
\begin{align}
\begin{split}
    \frac{\hGm(z)}{\hGm(0)} = \frac{\sinh k(z+h)}{\sinh kh} &+ k\int_{-h}^z \rmd \zeta\, q(\zeta)\frac{\sinh k(\zeta+h) \sinh k(z-\zeta)}{\sinh kh} \\
    &+ \frac{\sinh k(z+h)}{\sinh ^2 kh} k \int_{-h}^0 \rmd \zeta\, q(\zeta) \sinh k(\zeta+h) \sinh k\zeta    
\end{split}
\end{align}
into \eqref{eq:Dispersion_before_weak_curvature}, so we get the terms
\begin{align}
     \int_{-h}^0 \rmd z\, \Ug''(z) \frac{\Delta\Ug(z)}{\Omega(z)} \frac{\hGm(z)}{\hGm(0)} \frac{\sinh k (z+h)}{\cosh kh} = -2 k \int_{-h}^0 \rmd z\, q(z) \frac{\sinh^2 k(z+h)}{\sinh 2 k h} \Delta \Ug(z)
\end{align}
and
\begin{align}
        \int_{-h}^0 \rmd z\, \Ug'(z)\,\frac{d}{dz} \left(\frac{\hGm(z)}{\hGm(0)} \frac{\sinh k(z+h)}{\cosh kh} \right) &= 2 \omega_0 s \nonumber\\
        + 2 k\omega_0 s & \int_{-h}^0 \rmd z\, q(z) \frac{\sinh k(z+h) \sinh kz}{\sinh kh} \\
    + k^2 \int_{-h}^0 & \rmd z\, q(z) \frac{\sinh k(z+h)}{\sinh kh \cosh kh} \int_{z}^0\rmd \zeta \, \Ug(\zeta) \sinh k(2\zeta + h - z). \nonumber
\end{align}
All together, we get the approximate dispersion relation \eqref{eq:EL_dispersion} with the definitions of $\langle\Ug\rangle$, $s$, and $\Delta$ in \eqref{eq:effDopp}, \eqref{eq:s}, and \eqref{eq:Delta}, respectively.

\bibliographystyle{unsrtnat}
\bibliography{refs}

\end{document}